# Opening the Black Boxes in Data Flow Optimization

Fabian Hueske*[1] Mathias Peters†[2] Matthias J. Sax†[3] Astrid Rheinländer†[4]
Rico Bergmann†[5] Aljoscha Krettek*[6] Kostas Tzoumas*[7]
*Technische Universität Berlin, Germany †Humboldt-Universität zu Berlin, Germany
*[1,7]{fabian.hueske,kostas.tzoumas}@tu-berlin.de *[6]aljoscha.krettek@campus.tu-berlin.de
†[2,3,4,5]{mathias.peters,mjsax,rheinlae,bergmann}@informatik.hu-berlin.de

## ABSTRACT

Many systems for big data analytics employ a data flow abstraction to define parallel data processing tasks. In this setting, custom operations expressed as user-defined functions are very common. We address the problem of performing data flow optimization at this level of abstraction, where the semantics of operators are not known. Traditionally, query optimization is applied to queries with known algebraic semantics. In this work, we find that a handful of properties, rather than a full algebraic specification, suffice to establish reordering conditions for data processing operators. We show that these properties can be accurately estimated for black box operators by statically analyzing the general-purpose code of their user-defined functions.

We design and implement an optimizer for parallel data flows that does not assume knowledge of semantics or algebraic properties of operators. Our evaluation confirms that the optimizer can apply common rewritings such as selection reordering, bushy join-order enumeration, and limited forms of aggregation push-down, hence yielding similar rewriting power as modern relational DBMS optimizers. Moreover, it can optimize the operator order of non-relational data flows, a unique feature among today's systems.

## 1. INTRODUCTION

We are witnessing a data explosion in a variety of domains, including large-scale scientific data collection from various sensors, user-generated data, and data resulting from tracking human behavior online or otherwise. For example, the Large Hadron Collider at CERN generates around 15 petabytes per year [1], and the LSST telescope is expected to generate about 0.5 petabytes per month when it becomes operational [8]. Similar data volumes are expected to be created by next-generation DNA sequencing technologies [6]. It is now widely believed that a number of future scientific breakthroughs will be empowered by the ability to quickly analyze vast amounts of data. Similarly, the competitive advantage of many enterprises that operate on a web scale critically depends on drawing insights from huge data sets.

During the last years, it became clear that relational DBMSs could not cope with the scale and the nature of today's big data problems. This is due to a variety of reasons, including obsolete architectures [30], and trying to "fit" new problems to the relational model of programming. In 2004, Google reported their results on analyzing 100 terabytes of (mostly unstructured) data per day using their MapReduce framework [17], a number that grew to 20 petabytes per day in 2008 [18]. Partly motivated by these breakthroughs, new big data analysis systems have emerged to serve the aforementioned needs. Many of these systems such as Hyracks [11], Dryad [25], and our own Stratosphere system [7] adopt a *data flow* abstraction, where a data analysis program is specified as a directed acyclic graph (DAG) of smaller components that contain arbitrary user code. Even though some of these systems offer higher-level language interfaces [10, 12, 28, 31], supporting parallel user-defined functions (UDFs) is a fundamental requirement for these systems. Recently, commercial parallel DBMSs such as Aster Data and Greenplum have adopted MapReduce-style UDFs [2, 20] to explore a wider scope of applications.

The common challenge faced by these systems is to efficiently execute parallel data flows that embed UDFs. This entails parallelization, as well as reordering of operators. These two problems are highly coupled, as the optimal parallelization strategy depends on the operator order and vice versa. Traditional RDBMS optimizers support only UDFs that follow very strict templates such as scalar, aggregation, and table-generator UDFs. Due to these strict templates, the main challenge for RDBMS optimizers is not whether UDFs can be reordered but rather when it is beneficial. In contrast, MapReduce-style UDFs implement much less restrictive templates and hide their semantics inside general-purpose imperative code, a fact that poses new challenges for optimization. Conventional wisdom dictates that query optimization is possible at an abstraction layer where the semantics and the algebraic properties of operators are known. In this work, we build a query optimizer that does not require this assumption. Rather, our optimizer performs a fully automatic static code analysis pass over the UDFs, discovering a handful of properties that guarantee safe reorderings. We observe that a few properties, rather than knowledge of full semantics, are enough to enable many optimizations, including selection and join reordering, as well as limited forms of aggregation push-down.

The contributions of this paper are the following:

1. We introduce the problem of reordering data flow programs that consist of arbitrary imperative user-defined functions.
2. We formally establish the necessary conditions to reorder UDFs with a fixed signature (e. g., Map and Reduce) in a data flow.
3. We show how to derive the necessary knowledge for reordering via a static code analysis pass over the imperative UDF implementations.

4. We design and implement a query optimizer for this setting. In particular, we present a novel plan enumeration algorithm that does not use algebraic properties.
5. We implement the above concepts in the Stratosphere system [4], and conduct an extensive experimental study.
6. Our experimental results show that we can reproduce most reorderings done by traditional query optimizers in relational queries such as join and selection reordering and some forms of aggregation push-down. Further, our system can automatically find optimal plans for non-relational tasks *without* being informed a priori about the semantics of the operators.

While we present our optimizer in the context of the Stratosphere system, the results presented in this paper are applicable to a variety of parallel data flow systems that use imperative UDFs.

The remainder of this paper is organized as follows. Section 2 presents background material on Stratosphere's architecture, data model, and programming model. Section 3 introduces the problem by means of an example, and outlines the salient points of our solution. Section 4 delves into the details, and presents formal proofs for rewriting operators. Section 5 shows how to derive the information required by the optimizer using static code analysis. Section 6 presents the design of our query optimizer, including the plan enumeration algorithm. Section 7 presents our experimental study. Finally, Section 8 presents related work, and Section 9 concludes and offers research directions.

# 2. BACKGROUND: STRATOSPHERE

## 2.1 System Architecture

The Stratosphere system consists of two distinct components: The Nephele execution engine [7, 32], and the PACT compiler [7]. The user writes data analysis tasks in Java by providing first-order functions for a fixed set of second-order functions called Parallelization Contracts (PACTs, see Section 2.3 for details). The PACT compiler is responsible for translating the user-defined program into an efficient DAG data flow program, which is then deployed and executed by the Nephele engine. During compilation, the PACT compiler can exploit some declarative aspects of the PACT program in order to make cost-based decisions similar to a relational DBMS query optimizer, i. e., it decides on data shipping and local execution strategies for operators [7]. For example, the PACT compiler chooses between a partitioning, replication, or combined strategy for a parallel join (which is specified using the Match second-order function in the PACT programming model). The work described in this paper enables the PACT compiler to reorder operators in the data flow, in addition to choosing parallelization strategies.

## 2.2 Data Model

Stratosphere has recently migrated from a key-value pair data model to a record data model. The reasons for the new data model are twofold: First, it increases end-user productivity by allowing the programmer to work with more structured data rather than coalescing the data to a single value at every step of the data analysis program. Second, the new data model exposes more knowledge about the data analysis task to the compiler, making several new optimizations possible. For example, the optimizations presented in this paper would be rather limited if a simple key-value data model was used.

We define a *data set* as an unordered list of records, and denote it by $D = [r_1, \ldots, r_n]$. A record is an ordered tuple of values, $r = \langle v_1, \ldots, v_m \rangle$. The semantics of the values, including their type is left to the user-defined functions that manipulate them. We define two data sets $D_1, D_2$ as equal (denoted as $D_1 \equiv D_2$) when there exist two orderings of their records, such that $D_1 = [r_{11}, \ldots, r_{1n}]$, $D_2 = [r_{21}, \ldots, r_{2m}]$, $n = m$ and $\forall i = 1, \ldots, n : \ r_{1i} \equiv r_{2i}$. Two records $r_1 = \langle v_{11}, \ldots, v_{1n} \rangle$ and $r_2 = \langle v_{21}, \ldots, v_{2m} \rangle$ are equal ($r_i \equiv r_2$) iff $n = m$ and $\forall i = 1, \ldots, n : \ v_{1i} = v_{2i}$.

## 2.3 Programming Model

The PACT programming model [7] is a generalization of the MapReduce programming model [17]. A PACT program is a directed acyclic data flow composed of data sources, data sinks, and operators. An operator consists of a second-order function and an associated first-order user-defined function (UDF). In addition, some second-order functions require the specification of special (possibly composite) "key" fields. The first-order UDF can emit an arbitrary number of output records per invocation, possibly with modified value types. The second-order function defines how the input data set is partitioned into *groups* and applies the first-order function to each group independently. Hence, groups are processed in a data-parallel fashion possibly on different nodes without incurring communication overhead. Thereby, the type of the second-order function defines the parallelization opportunities for a given operator.

There are currently five second-order functions (called PACTs) implemented in Stratosphere: Map, Reduce, Cross, Match, and CoGroup (see Figure 1). The Map function dictates that every input record forms an individual group. The Reduce function dictates that a group exists for every unique value of the key attribute in the input data set, and contains all records with the particular key value. The Cross, Match, and CoGroup second-order functions are used to define binary operators. The Cross function forms a group from every pair of records in its two inputs, similarly to forming a distributed Cartesian product of two sets. The Match function forms a group from every pair of records in its two inputs, only if the records have the same value for the key attribute. Match is therefore similar to an equi-join. Finally, the CoGroup function forms a group for every value of the key attribute (from the domains of both inputs), and places each record in the appropriate group depending on the key value of the record.

More formally, assume two input data sets $R = [r_1, \ldots, r_N]$ and $S = [s_1, \ldots, s_M]$. The Map PACT is defined as

$$\texttt{Map} : R \times f \rightarrow [f(r_1), \ldots, f(r_N)]$$

where $f$ is the user-defined first-order function of Map. The Reduce function is defined as

$$\texttt{Reduce} : R \times f \times \mathsf{K} \rightarrow [f(r_1^{\mathsf{k}_1}, \ldots, r_{n_1}^{\mathsf{k}_1}), \ldots, f(r_1^{\mathsf{k}_l}, \ldots, r_{n_l}^{\mathsf{k}_l})]$$

where $\mathsf{K}$ is a set of attributes of $R$ called the key, the active domain of $\mathsf{K}$ in $R$ is $\{\mathsf{k}_1, \ldots, \mathsf{k}_l\}$, and for record $r^{\mathsf{k}}$ it holds that $r.\mathsf{K} = \mathsf{k}$. Note that the UDF $f$ of a Reduce function operates on a list of input records. The Cross and Match functions are defined as

$$\texttt{Cross} : R \times S \times f \rightarrow [f(r_1, s_1), f(r_1, s_2), \ldots, f(r_N, s_M)]$$

$$\texttt{Match} : R \times S \times \mathsf{K} \times \mathsf{F} \times f \rightarrow [\{f(r, s) | r.\mathsf{K} = s.\mathsf{F}\}]$$

where $\mathsf{K}$ and $\mathsf{F}$ are the keys of the Match function for $R$ and $S$ respectively. The CoGroup function is defined as

$$\texttt{CoGroup} : R \times S \times \mathsf{K} \times \mathsf{F} \times f \rightarrow$$
$$[f(r_1^{\mathsf{v}_1}, \ldots, r_{n_1}^{\mathsf{v}_1}, s_1^{\mathsf{v}_1}, \ldots, s_{m_1}^{\mathsf{v}_1}), \ldots, f(r_1^{\mathsf{v}_l}, \ldots, r_{n_l}^{\mathsf{v}_l}, s_1^{\mathsf{v}_l}, \ldots, s_{m_l}^{\mathsf{v}_l})]$$

where the combined active domain of $\mathsf{K}$ and $\mathsf{F}$ is $\{\mathsf{v}_1, \ldots, \mathsf{v}_l\}$.

We distinguish between PACTs whose UDF is called with exactly one record per input (Map, Match, and Cross) as argument and PACTs whose UDF is called with a list of records per input (Reduce and CoGroup). We call the former *record-at-a-time (RAT)* operators, and the latter *key-at-a-time (KAT)* operators. For the latter, we refer to all input records of data set $D$ with a specific key value $\mathsf{k}$ as a *key group* $D_{\mathsf{k}}$.

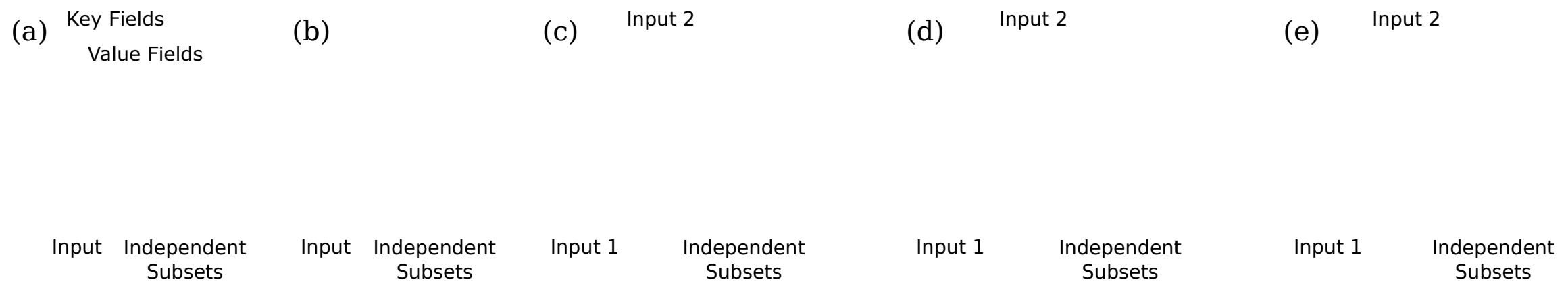

**Figure 1: (a) Map, (b) Reduce, (c) Cross, (d) Match, and (e) CoGroup second-order functions.**

# 3. A REORDERING EXAMPLE

We address the concrete problem of optimizing PACT programs, in which the algebraic properties of first-order functions are not known. Our solution proceeds in three steps: First, in Section 4, we establish the necessary conditions to reorder PACT operators. At this stage, we treat the UDFs of operators as black boxes. Our key insight is that a few properties, rather than full semantics, suffice to establish many reordering conditions. Next, in Section 5, we show how to safely approximate these properties, by "opening" the black box operators via a static code analysis pass over their code. Finally, in Section 6, we show how to enumerate plans when the concept of algebraic expressions does not apply. We first demonstrate the salient points of our complete solution with an example.

Assume a PACT program $P$ that consists of three Map operators with first-order functions $f_1$, $f_2$, and $f_3$ interconnected as follows:

$$P : I \rightarrow \mathtt{Map}_1 \rightarrow \mathtt{Map}_2 \rightarrow \mathtt{Map}_3 \rightarrow O$$

The input data set $I$ contains two integer attributes $\langle A, B\rangle$. The first function $f_1$ replaces $B$ with $|B|$. The second function $f_2$ emits all records for which $A \geq 0$ and filters the rest of the records, and the third function $f_3$ replaces $A$ with the sum $A + B$. For example, with input record $i = \langle 2, -3\rangle$, the data flow is

$$\langle 2, -3\rangle \rightarrow f_1 \rightarrow \langle 2, 3\rangle \rightarrow f_2 \rightarrow \langle 2, 3\rangle \rightarrow f_3 \rightarrow \langle 5, 3\rangle$$

while with input record $i' = \langle -2, -3\rangle$ the data flow is

$$\langle -2, -3\rangle \rightarrow f_1 \rightarrow \langle -2, 3\rangle \rightarrow f_2 \rightarrow \perp \rightarrow f_3 \rightarrow \perp$$

where $\perp$ represents the empty list.

Consider now the alternative plan $P'$ where the order of $\mathtt{Map}_2$ and $\mathtt{Map}_1$ is inverse:

$$P' : I \rightarrow \mathtt{Map}_2 \rightarrow \mathtt{Map}_1 \rightarrow \mathtt{Map}_3 \rightarrow O$$

The data flow for records $i$ and $i'$ is

$$\langle 2, -3\rangle \rightarrow f_2 \rightarrow \langle 2, -3\rangle \rightarrow f_1 \rightarrow \langle 2, 3\rangle \rightarrow f_3 \rightarrow \langle 5, 3\rangle$$
$$\langle -2, -3\rangle \rightarrow f_2 \rightarrow \perp \rightarrow f_1 \rightarrow \perp \rightarrow f_3 \rightarrow \perp$$

Observe that the order of $\mathtt{Map}_1$ and $\mathtt{Map}_2$ does not influence the output data set $O$. Therefore, for input $I = [i, i']$, these two operators can be safely reordered. In fact, if $f_2$ filters a significant portion of the records in $I$, this reordering is desirable. On the other hand, $f_1$ and $f_3$ cannot be further reordered without changing the result:

$$\langle 2, -3\rangle \rightarrow f_2 \rightarrow \langle 2, -3\rangle \rightarrow f_3 \rightarrow \langle -1, -3\rangle \rightarrow f_1 \rightarrow \langle -1, 3\rangle$$

We generalize this concept in a safe manner without knowing the semantics of the operators. Our key insight is that *reasoning about the "conflicts" in the data flow suffices to establish reordering conditions*. For example, we do not need to know whether $f_3$ computes $A + B$ or $A \cdot B$. We only need to know that $f_3$ replaces the first field of its input record with a new value, which conflicts with $f_2$ using the first field of its input record to potentially filter some records. We can therefore establish that these operators "conflict" on $A$, and cannot be reordered. This holds only if *the execution path of a UDF is uniquely determined by its input data*, i. e., communication between functions except via the explicitly defined data channels of the data flow program (e. g., shared memory or other forms of communication) is prohibited. We assume this restriction throughout this paper.

We define a *read set* $\mathsf{R}_f$, and a *write set* $\mathsf{W}_f$ for each operator with respect to its UDF $f$. These sets are defined over *attributes* that need to be extracted from the plan. In our example plan, we have two attributes $A, B$, that form the so-called *global record* $\mathsf{A} = \{A, B\}$. The read set of an operator contains all attributes that *might influence the operator's output*. The write set of an operator contains all attributes *whose values change with an application of the operator*. We formalize these concepts in Section 4. Two operators "conflict" on an attribute if the attribute is contained in both operators' write sets, or in one operator's read set and the other's write set. For example, operator $f_1$ has $\mathsf{R}_{f_1} = \{B\}$, and $\mathsf{W}_{f_1} = \{B\}$, and operator $f_2$ has $\mathsf{R}_{f_2} = \{A\}$, and $\mathsf{W}_{f_2} = \emptyset$. These operators do not conflict, and can therefore be reordered.

The next challenge we address is how to derive read and write sets among other necessary properties. In Section 5 we present an algorithm that *estimates* these properties using a static code analysis (SCA) pass over the code of the first-order functions. Assume the code of the three example first-order functions shown below in the form of 3-address code [5] where the UDFs access fields $A$ and $B$ by their positions (0 and 1 respectively) in the input record:

```
20: f2(InputRecord $ir)
21: $a:=getField($ir,0)
22: if($a<0) goto 25
23: $or:=copy($ir)
24: emit($or)
25: return
```

```
10: f1(InputRecord $ir)
11: $b:=getField($ir,1)
12: $or:=copy($ir)
13: if ($b>=0) goto 16
14: $b:=-$b
15: setField($or,1,$b)
16: emit($or)
17: return
```

```
30: f3(InputRecord $ir)
31: $a:=getField($ir,0)
32: $b:=getField($ir,1)
33: $sum:=$a+$b
34: $or=copy($ir)
35: setField($or,0,$sum)
36: emit($or)
37: return
```

The instructions with labels 10 to 17 are the code of function $f_1$, with labels 20 to 25 of $f_2$, and with labels 30 to 37 of $f_3$. Consider for example the code of function $f_2$. Recall that $f_2$ filters records with negative values for attribute $A$. We can automatically detect that $A \in \mathsf{R}_{f_1}$ by collecting all `getField` statements (in this case instruction 21), and determining whether the temporary variables introduced (in this case `$a`) are used in the function's code. In our example, instruction 22 uses the value of `$a` in a condition, so we conclude that field 0 of the input record is part of the read set. In the same way, we can detect that $A \in \mathsf{W}_{f_3}$ by looking at instruction 35, which potentially changes the value of field 0. We can thus conclude that $f_2$ and $f_3$ conflict on field 0, and cannot be re-ordered. This estimation is conservative, but safe. It results in a set of reorderings that all produce the same query result, but it might miss valid reorderings. For example, assume that the input data set $I$ contains only values with $A \geq 0$. Then, instructions 22 and 23 of function $f_2$ will never be executed, and in fact, $f_2$ and $f_3$ can be reordered. However, this is something that cannot be detected by static code analysis, and this reordering will be prohibited by our system.

# 4. CONDITIONS FOR REORDERING

## 4.1 Definitions

The user-code of operators accesses record attributes by static field indices. However, the reordering of two operators can cause changes of the mapping of field indices to attributes. Since the user-code assumes the original mapping, it is essential to avoid that attributes are accessed by wrong indices in order to preserve the original semantics of the data flow. For this purpose, we define the *global record* as a collection of every attribute that is accessed by any operator in the execution plan. Thus, the global record includes every attribute of the input data sets as well as the attributes that are created by operators at some stage of the execution plan.

**Definition 1.** The global record $\mathsf{A}$ is a unique naming of all base and intermediate attributes in the data flow. In addition, we define a *redirection map* $\alpha(D, n)$, which maps every field index $n \in \mathbb{N}$ of every data set $D$ (base or intermediate) to the corresponding entry in the global record $\mathsf{A}$.

Next, we formally define the read and write sets. Denote by $\mathsf{D}$ the attributes of data set $D$, and by $\#\mathsf{D}$ the number of attributes of $D$. The write set $\mathsf{W}_f$ of an operator with first-order function $f$ contains the attributes whose value might change after applying $f$.

**Definition 2.** An attribute $A$ belongs to the write set $\mathsf{W}_f$ of an operator with UDF $f$, input $I$, and output $O$ iff:

(1) $A = \alpha(O, m),\ m > \#\mathsf{I}$, or

(2) $A = \alpha(I, n),\ \exists i \in \textit{Instances}(I) : \exists o_i \in f(i) : o_i[n] \neq i[n]$

The definition captures the fact that an attribute is in $\mathsf{W}_f$ if it is either newly created by $f$ (case 1 of the definition), or that there exists at least one record in the data set with a different value of this attribute after $f$ is applied (case 2 of the definition). The above definition can be extended for UDFs that operate on multiple records. The read set $\mathsf{R}_f$ of an operator with user function $f$ contains the attributes that might influence the operators's output.

**Definition 3.** An attribute $A = \alpha(I, n)$ belongs to the read set $\mathsf{R}_f$ of an operator with UDF $f$, input $I$, and output $O$ iff:

$\exists i_1, i_2 \in \textit{Instances}(I), \forall m \neq n$: $i_1[n] \neq i_2[n] \wedge i_1[m] = i_2[m]$

(1) $(|f(i_1)| \neq |f(i_2)|)$, or

(2) $(\exists o_1 \in f(i_1), o_2 \in f(i_2), k \neq n : o_1[k] \neq o_2[k])$

The definition captures the fact that an attribute $A$ can influence $f$'s output if a change on $A$'s value only may produce a different output. Note that key attributes of KAT operators are always included to the read set because they directly influence the operator's result. Note that the above definitions do not use the semantics of the functions. Section 5 discusses how to approximate these sets using static code analysis of the UDFs.

Finally, we define two conditions that are necessary for reordering of operators in most cases:

**Definition 4.** Two operators with UDFs $f_1$, $f_2$ satisfy the *read-only conflict (ROC) condition* iff $\mathsf{R}_{f_1} \cap \mathsf{W}_{f_2} = \mathsf{W}_{f_1} \cap \mathsf{R}_{f_2} = \mathsf{W}_{f_1} \cap \mathsf{W}_{f_2} = \emptyset$.

The ROC condition captures the fact that a UDF does not update or use attributes that another UDF updates. The ROC condition is necessary for all reorderings described in this paper. To reorder KAT operators, we additionally need the condition that key groups are preserved:

**Definition 5.** An operator with UDF $f$ satisfies the *key group preservation (KGP) condition* for an attribute set $\mathsf{K} \subset \mathsf{A}$ iff (1) $\forall r \in I$ : $|f(r)| = 1$, or (2) $|f(r)| < 1$, and $\exists \mathsf{F}, \mathsf{F} \subset \mathsf{K} : \forall r, r' \in I$ : $\pi_{\mathsf{F}}(r) = \pi_{\mathsf{F}}(r') \Rightarrow |f(r)| = |f(r')|$

The projection $\pi$ of a record on a set of attributes is defined as usual. The above definition can be extended for KAT operators. The KGP condition states that function $f$, when applied to a set of records $I_k$ with the same value for $\mathsf{K}$, either emits or filters all these records.

## 4.2 Reordering MapReduce Programs

### 4.2.1 Reordering Map Operators

In Section 3 we outlined why two Map operators that satisfy the ROC condition can be reordered without changing the query result. We now prove this statement formally.

**Theorem 1.** Two Map operators can be reordered if their first-order functions satisfy the ROC condition.

PROOF. Assume the two plans

$$P : I \to \mathtt{Map}_f \to S \to \mathtt{Map}_g \to O$$
$$P' : I \to \mathtt{Map}_g \to S' \to \mathtt{Map}_f \to O'$$

We prove that $O \equiv O'$. Assume a record $i \in I$, and let $O_i = \mathtt{Map}_g(\mathtt{Map}_f([i]))$, $O'_i = \mathtt{Map}_f(\mathtt{Map}_g([i]))$, $S_i = \mathtt{Map}_f([i]) = f(i)$, and $S'_i = \mathtt{Map}_g([i]) = g(i)$. It suffices to prove $\forall i \in I : O_i \equiv O'_i$. We first observe that if the ROC condition holds, the global record can be partitioned as $\mathsf{A} = \overline{\mathsf{W}} \cup (\mathsf{W}_f \dot{\cup} \mathsf{W}_g)$, where $A \dot{\cup} B$ additionally implies that $A \cap B = \emptyset$. We define $\pi_{\mathsf{F}}(r)$ as the projection of record $r$ to attribute subset $\mathsf{F}$.

First, we prove that an invocation of $f$ and $g$ produces the same result cardinality in both plans: $|f(i)| = |f(s'_j)| = k$ where $s'_j \in S'_i$, and $|g(i)| = |g(s_i)| = l$ where $s_i \in S_i$. Records $s'_j \in S'_i$ are produced by applying $g$ to $i$. Recall that $g$ can only change $\mathsf{W}_g$ attributes, therefore $\pi_{\overline{\mathsf{W}} \cup \mathsf{W}_f}(s'_j) = \pi_{\overline{\mathsf{W}} \cup \mathsf{W}_f}(i)$. Observe that the execution path of $f$ depends only on the values of attributes in $\overline{\mathsf{W}} \cup \mathsf{W}_f$. Therefore, $f$ follows the same execution path for $s'_j$ and $i$, and the cardinality of its output is the same: $\forall s'_j : |f(i)| = |f(s'_j)| = k$. We can similarly prove $|g(i)| = |g(s_i)| = l$. This allows us to decompose plan $P$ for input $i$ as

$$P_1 : i \to f \to [s_i | i = 1, \ldots, k]$$
$$P_2 : s_i \to g \to [o_{ij} | j = 1, \ldots, l]\ \forall i = 1, \ldots, k$$

and plan $P'$ as

$$P'_1 : i \to g \to [s'_j | j = 1, \ldots, l]$$
$$P'_2 : s'_j \to f \to [o'_{ji} | i = 1, \ldots, k]\ \forall j = 1, \ldots, l$$

We will now prove that $\forall i = 1, \ldots, k,\ \forall j = 1, \ldots, l : o_{ij} = o'_{ji}$. We observe that $\pi_{\overline{\mathsf{W}}}(o_{ij}) = \pi_{\overline{\mathsf{W}}}(o'_{ji})$ since attributes in $\overline{\mathsf{W}}$ are not changed by either $f$ or $g$. Therefore, it suffices to prove that (1) $\pi_{\mathsf{W}_f}(o_{ij}) = \pi_{\mathsf{W}_f}(o'_{ji})$, and (2) $\pi_{\mathsf{W}_g}(o_{ij}) = \pi_{\mathsf{W}_g}(o'_{ji})$. The proofs for the two cases are completely symmetric. We proceed to prove case (1).

From sub-plan $P_2$ we observe that records $o_{ij}$ are produced by applying $g$ to records $s_i$. Therefore, they have the same values for all attributes that $g$ does not change: $\pi_{\overline{\mathsf{W}} \cup \mathsf{W}_f}(o_{ij}) = \pi_{\overline{\mathsf{W}} \cup \mathsf{W}_f}(s_i)$. It suffices thus to prove $\pi_{\mathsf{W}_f}(o'_{ji}) = \pi_{\mathsf{W}_f}(s_i)$. Consider sub-plans $P_1$ and $P'_2$ that show the application of $f$ to $i$ and $s'_j$ respectively. First, observe that $s'_j$ comes from applying $g$ to $i$, therefore $\pi_{\mathsf{W}_f}(s'_j) = \pi_{\mathsf{W}_f}(i)$. The execution path of $f$ depends only on values of attributes in $\overline{\mathsf{W}} \cup \mathsf{W}_f$. Since $\pi_{\mathsf{W}_f}(s'_j) = \pi_{\mathsf{W}_f}(i)$, the execution of $f$ in sub-plans $P_1$ and $P'_2$ will follow the same execution path. Therefore, the changes applied to $i$ will be the same as the changes applied to $s'_j$. Therefore, $\pi_{\mathsf{W}_f}(o'_{ji}) = \pi_{\mathsf{W}_f}(s_i)$. □

### 4.2.2 *Reordering Map and Reduce Operators*

Recall that unlike the MapReduce model, the PACT model allows arbitrary data flows containing Map and Reduce (among other) operators. Assume the plan

$$P : I \rightarrow \texttt{Map}_f \rightarrow S \rightarrow \texttt{Reduce}_g \rightarrow O$$

with input $I$ having two attributes $\langle A, B \rangle$. UDF $f$ emits all input records with odd values both of $A$ and $B$. UDF $g$ calculates the sum of $B$ using $A$ as key, and appends the sum as a new attribute $C$ to all of its input records. Note that the ROC condition holds. Consider the input data set in the following example application of the plan:

$$\begin{bmatrix} \langle 1,1\rangle, \langle 1,2\rangle, \\ \langle 2,1\rangle, \langle 2,2\rangle \end{bmatrix} \rightarrow \texttt{Map}_f \rightarrow [\langle 1,1\rangle] \rightarrow \texttt{Reduce}_g \rightarrow [\langle 1,1,1\rangle]$$

and the execution if the operators are reordered

$$\begin{bmatrix} \langle 1,1\rangle, \langle 1,2\rangle, \\ \langle 2,1\rangle, \langle 2,2\rangle \end{bmatrix} \rightarrow \texttt{Reduce}_g \rightarrow \begin{bmatrix} \langle 1,1,3\rangle, \\ \langle 1,2,3\rangle, \\ \langle 2,1,3\rangle, \\ \langle 2,2,3\rangle \end{bmatrix} \rightarrow \texttt{Map}_f \rightarrow [\langle 1,1,3\rangle]$$

The ROC condition alone cannot guarantee the reordering of a Map and a Reduce operator. The reason is that the key groups of the Reduce operator in the two plans do not have the same cardinality, and thus result in a different value for attribute $C$. This would not be a problem if the Map operator either eliminated whole key groups, or left them intact. Note that if Map also emitted multiple records per call, the cardinality of the key groups would change. Therefore, we need the KGP condition to hold as well.

**Theorem 2.** A Map operator with UDF $f$ and a Reduce operator with UDF $g$ can be reordered if the ROC condition holds for $f, g$, and the KGP condition holds for $f$ and the key $\mathsf{K}$ of the Reduce operator.

PROOF. Consider the two pipelines:

$$P : I \rightarrow \texttt{Map}_f \rightarrow S \rightarrow \texttt{Reduce}_g \rightarrow O$$
$$P' : I \rightarrow \texttt{Reduce}_g \rightarrow S' \rightarrow \texttt{Map}_f \rightarrow O'$$

As before, we prove that $O \equiv O'$. Let $I = \cup_{\mathsf{k}} I_{\mathsf{k}}$, where $I_{\mathsf{k}}$ is the key group with key value $\mathsf{k}$ and the plans

$$P : I_{\mathsf{k}} \rightarrow \texttt{Map}_f \rightarrow S_{\mathsf{k}} \rightarrow \texttt{Reduce}_g \rightarrow O_{\mathsf{k}}$$
$$P' : I_{\mathsf{k}} \rightarrow \texttt{Reduce}_g \rightarrow S'_{\mathsf{k}} \rightarrow \texttt{Map}_f \rightarrow O'_{\mathsf{k}}$$

It suffices to prove that $O_{\mathsf{k}} \equiv O'_{\mathsf{k}}$. Observe that if the KGP condition holds, $|S_{\mathsf{k}}| = |I_{\mathsf{k}}|$, or $|S_{\mathsf{k}}| = 0$. If $|S_{\mathsf{k}}| = 0$, then $\texttt{Map}_f$ will also filter all records from $S'_k$ in $P'$, and trivially $O_{\mathsf{k}} \equiv O'_{\mathsf{k}} = \perp$. Assume that $|I_{\mathsf{k}}| = |S_{\mathsf{k}}| = k$, and $|O_{\mathsf{k}}| = l$. Since the Reduce UDF treats $I_{\mathsf{k}}$ in $P'$ in the same way as $S_{\mathsf{k}}$ in $P$ (because $|I_{\mathsf{k}}| = |S_{\mathsf{k}}|$ and the ROC condition holds), and the Map UDF emits exactly one record per input, it holds that $|S'_{\mathsf{k}}| = |O'_{\mathsf{k}}| = l$. Therefore, we can decompose plan $P$ as

$$P_1 : \quad \forall i \in [i_1, \ldots, i_k],\ i \rightarrow f \rightarrow s,\ s \in [s_1, \ldots, s_k]$$
$$P_2 : \quad [s_1, \ldots, s_k] \rightarrow g \rightarrow [o_1, \ldots, o_l]$$

and plan $P'$ as

$$P'_1 : \quad [i_1, \ldots, i_k] \rightarrow g \rightarrow [s'_1, \ldots, s'_l]$$
$$P'_2 : \forall s' \in [s'_1, \ldots, s'_l],\ s' \rightarrow f \rightarrow o',\ o' \in [o'_1, \ldots, o'_l]$$

We now prove that $\forall j, j = 1, \ldots, l : o_j = o'_j$. Due to the ROC condition it suffices to prove (1) $\pi_{\mathsf{W}_f}(o_j) = \pi_{\mathsf{W}_f}(o'_j)$, and (2) $\pi_{\mathsf{W}_g}(o_j) = \pi_{\mathsf{W}_g}(o'_j)$.

We proceed to prove case (1). Case (2) is proven similarly. From sub-plan $P_2$, and record $o_j$, there is a record $s_x$ with the same attribute values for $\mathsf{W}_f$: $\forall j, j = 1, \ldots, l\ \exists x, x = 1, \ldots, k : \pi_{\mathsf{W}_f}(o_j) = \pi_{\mathsf{W}_f}(s_x)$. Note that Reduce may "consolidate" multiple records into one, or produce multiple records per input record. However, due to the ROC condition, attributes in $\mathsf{W}_f$ must be preserved. Similarly, from $P'_1$ we have $\forall j, j = 1, \ldots, l\ \exists y, y = 1, \ldots, k : \pi_{\mathsf{R}_f}(s'_j) = \pi_{\mathsf{R}_f}(i_y)$ (due to the ROC condition, attributes in $\mathsf{R}_f$ are preserved as well). Using the same arguments as in the proof of Theorem 1, we know that $g$ follows the same execution path in sub-plans $P_2$ and $P'_1$.

Therefore, $f$ follows the same execution path for records $s'_j$ and $i_x$, so the result records of applying $f$ to these records will also share the same values for $\mathsf{W}_f$ attributes: $\pi_{\mathsf{W}_f}(o'_j) = \pi_{\mathsf{W}_f}(s_x) \Rightarrow \pi_{\mathsf{W}_f}(o'_j) = \pi_{\mathsf{W}_f}(o_j)$. □

The condition for reordering two Reduce operators are the ROC condition and the KGP condition for both UDF-key pairs. The proof proceeds similarly.

## 4.3 Reordering Binary Second-Order Functions

### 4.3.1 *Record-at-a-time Operators*

We first cover plans with RAT operators that are constructed using the Cross, Match, and Map PACTs. Assume a Cross operator with UDF $f$ and inputs $R, S$. The operator applies $f$ to every pair $(r, s) \in R \times S$. Here, the Cartesian product $R \times S$ of two data sets $R = [r_1, \ldots, r_n]$, $S = [s_1, \ldots, s_m]$ is defined as a data set $R \times S = [r_i | s_j : i = 1 \ldots, n, j = 1, \ldots, m]$ where $r|s$ is the concatenation of records $r$ and $s$. The attribute set of $R \times S$ is the union of the attribute sets of $R$ and $S$ with a proper renaming (e. g., each attribute is prefixed by the data set name it belongs to).

We observe that we can *conceptually* transform a Cross operator to a Map operator with the same UDF over the Cartesian product:

$$\texttt{Cross}_f(R, S) \equiv \texttt{Map}_f(R \times S)$$

We can similarly transform a Match operator with UDF $f$ to a Map operator with UDF $f'$ over the Cartesian product:

$$\texttt{Match}_f(R, S) \equiv \texttt{Map}_{f'}(R \times S)$$

The difference here is that we need to change the UDF $f$ in order to incorporate the implicit equi-join performed by the Match second-order function. Assume that the join keys are attributes $R.A, S.B$. We substitute $f$ with

$$f'(r|s) = \texttt{if}\ (R.A = S.B)\ \texttt{then}\ f(r, s)\ \texttt{else}\ \perp$$

We stress that this is a conceptual transformation in order to establish reordering conditions; all optimizations described in this paper are non-intrusive. This transformation simply means that the attributes used as keys for the Match operator are added to the read set $\mathsf{R}_f$ of the operator.

Using the above transformations, plans that contain Match, Cross, and Map operators are equivalent to plans that contain only Map operators and Cartesian products. Therefore, it only remains to establish when the latter two can be reordered:

**Theorem 3.** A Map operator with UDF $f$ and a Cartesian product operator $R \times S$ can be reordered as

$$\texttt{Map}_f(R \times S) \equiv \texttt{Map}_f(R) \times S$$

iff $(\mathsf{R}_f \cup \mathsf{W}_f) \cap \mathsf{S} = \emptyset$, where $\mathsf{S}$ is the attribute set of $S$. The case of pushing the operator to the other side of the Cartesian product is symmetric.

The proof follows directly from the fact that $(\mathsf{R}_f \cup \mathsf{W}_f) \cap \mathsf{S} = \emptyset \Rightarrow f(r|s) = f(r)|s$.

It is straightforward to construct the conditions that allow reordering for Match, Cross, and Map operators using Theorems 1 and 3. We now show the proof for reordering two Match operators with first order functions $f, g$, and key attributes $\mathsf{K}_f, \mathsf{K}_g$ as a series of transformations:

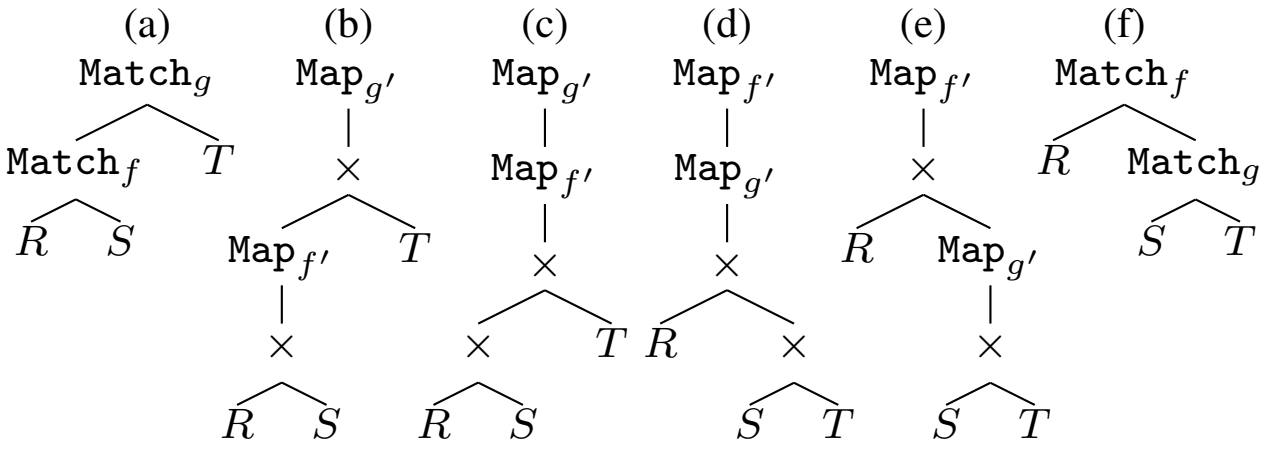

Step (a) $\rightarrow$ (b) substitutes the Match operators with their Map and Cartesian product equivalents. Step (b) $\rightarrow$ (c) reorders $\texttt{Map}_{f'}$ with the Cartesian product with $T$. For plans (b) and (c) to be equivalent it is necessary that $f'$ does not use attributes of $T$ ($(\mathsf{R}_{f'} \cup \mathsf{W}_{f'}) \cap \mathsf{T} = \emptyset$). Step (c) $\rightarrow$ (d) makes use of the conditions of Theorem 1 (namely the ROC condition on UDFs $f', g'$) to reorder the two Map operators, and reorders the two Cartesian products using the normal associativity rule. Step (d) $\rightarrow$ (e) pushes $\texttt{Map}_{g'}$ under the Cartesian product, requiring the condition $(\mathsf{R}_{g'} \cup \mathsf{W}_{g'}) \cap \mathsf{R} = \emptyset$. Finally, step (e) $\rightarrow$ (f) reconstructs the Match operators. By collecting the conditions needed by the series of transformations, we arrive at the conditions to reorder two Match operators.

**Lemma 1.** Two Match operators with UDFs $f, g$ and key sets $\mathsf{K}_f \subset \mathsf{R} \cup \mathsf{S}, \mathsf{K}_g \subset \mathsf{S} \cup \mathsf{T}$ can be reordered iff the ROC condition holds for $f', g'$, $(\mathsf{R}_{f'} \cup \mathsf{W}_f) \cap \mathsf{T} = \emptyset$, and $(\mathsf{R}_{g'} \cup \mathsf{W}_g) \cap \mathsf{R} = \emptyset$ where $\mathsf{R}_{f'} = \mathsf{R}_f \cup \mathsf{K}_f$, and $\mathsf{R}_{g'} = \mathsf{R}_g \cup \mathsf{K}_g$.

By repeating the same process for each pair of Match, Cross, and Map, we establish similar conditions for all combinations of these operators.

### 4.3.2 *Key-at-a-time Operators*

Incorporating KAT operators (Reduce and CoGroup) requires stricter conditions, since groups must be preserved. We first show how to reorder a Reduce operator with a Cartesian product.

**Theorem 4.** A Reduce operator with UDF $g$ and key $\mathsf{K} \cup \mathsf{R}$ and a Cartesian product operator $R \times S$ can be reordered as

$$\texttt{Reduce}_{g,\mathsf{K} \cup \mathsf{R}}(R \times S) \equiv R \times \texttt{Reduce}_{g,\mathsf{K}}(S)$$

iff $(\mathsf{R}_g \cup \mathsf{W}_g) \cap \mathsf{R} = \emptyset$.

PROOF. Assume the data sets $R = [r_i : i = 1 \ldots, n]$, $S = [s_i : i = 1 \ldots, m]$. The key of the Reduce operator $\mathsf{K} \cup \mathsf{R}$ includes all attributes of data set $R$. Note that $\mathsf{K} \subset \mathsf{S}$. Every record of the Cartesian product can be written as $r_i|k_j|s'_k$, where $k_j$ is the part of the $S$ record with attributes $\mathsf{K}$, and $s'_k$ is the part of an $S$ record with non-key attributes. Every record $r_i|k_j|s'_k$ of the Cartesian product belongs to the same Reduce group $G_{ij}$, determined by $r_i$ and $k_j$ only. The output of the plan is $[g(G_{ij}), G_{ij} = \{r_i|k_j|s'_k\}]$. Assume that $g$ does not use any attribute of $R$ for any purpose other than grouping its input data set. Then, it is safe to "push" $\texttt{Reduce}_g$ to the data set $S$ and remove the R part of the Reduce key. This will produce groups $G_j = \{k_j|s'_k\}$, and the output of the reduce will be $[g(G_j)]$. By performing the Cartesian product of these groups with $R$, we get the set of records $r_i|g(G_j)$. If the Reduce UDF $g$ simply emits the $R$ attributes unchanged, we have $r_i|g(k_j|s_k) = g(r_i|k_j|s_k)$. □

Using the above transformation, we can, in principle, reorder Reduce with Match and Cross operators by transforming the latter to Map operators over Cartesian products. It is not very often that the Reduce key includes all attributes of a data set. However, we can consider special cases where it is safe to *add* the R attributes to the Reduce key without changing the result. One case is when $|R| = 1$. This appears quite often in practice when implementing SQL queries with correlated subqueries that return a single tuple. More interestingly, using Theorem 4 as basis, we can arrive at a Match-Reduce transformation similar to the invariant grouping transformation in relational DBMSs [13]. Assume the plan

$$\text{(a) } \texttt{Reduce}_{g,\mathsf{F}}(\texttt{Match}_{f,R.\mathsf{K}=S.\mathsf{F}}(R, S))$$

where the Match keys are $\mathsf{K} \subset \mathsf{R}$, $\mathsf{F} \subset \mathsf{S}$, and the Reduce key is a superset of $\mathsf{F}$. Assume that $\mathsf{F}$ is a foreign key to $\mathsf{K}$. Then, in every record received by the Reduce operator, the $\mathsf{F}$ part uniquely determines all $R$ attributes. We can therefore add R to the key of the Reduce operator without changing the Reduce groups, and apply Theorem 4 to push the Reduce under the Match. As always, the ROC and KGP conditions must hold in order to reorder the Reduce and Map UDFs. The transformation steps taking plan (a) above as the starting point are shown below.

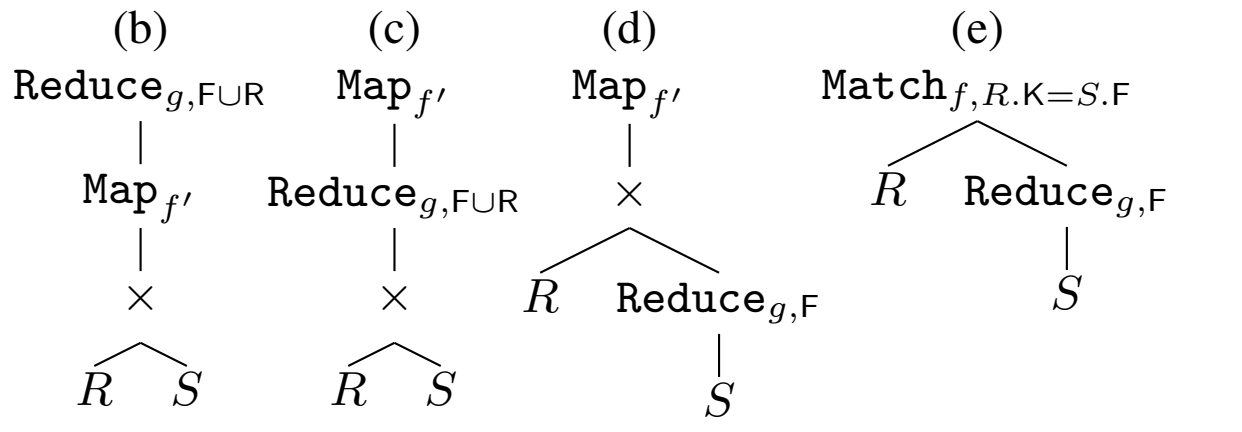

The last step is to incorporate CoGroup operators. We note that a CoGroup operator can be conceptually transformed to a Reduce operator over the *tagged union* $R \cup_T S$ of its inputs $R, S$:

$$\texttt{CoGroup}_g(R, S) \equiv \texttt{Reduce}_{g'}(R \cup_T S)$$

The tagged union of two data sets $R$ and $S$ is simply the data set $R$ followed by the data set $S$, where each record has an additional *lineage* attribute $l$, which tracks the data set that the record originates from. The CoGroup UDF $g$ is properly annotated to distinguish between data sets based on the lineage attribute, yielding the Reduce UDF $g'$.

Map and Reduce operators can be pushed down under the tagged union $R \cup_T S$ if their UDFs operate only on one of the tagged union's inputs. This can be properly detected using the lineage attribute $l$. For example, assume that we want to push a Map operator with UDF $f$ under the tagged union $R \cup_T S$, and that $f$ uses only $R$ attributes. We can define a UDF $f_R$ as

$$f_R(r) = \begin{cases} f(r) \text{ if } r.l = R \\ r \text{ otherwise.} \end{cases}$$

thus forcing the Map UDF $f$ to ignore $S$ records. This transformation yields

$$\texttt{Map}_{f_R}(R \cup_T S) \equiv \texttt{Map}_f(R) \cup_T S$$

and allows the following series of transformations that show how a Map operator can be reordered with a CoGroup operator.

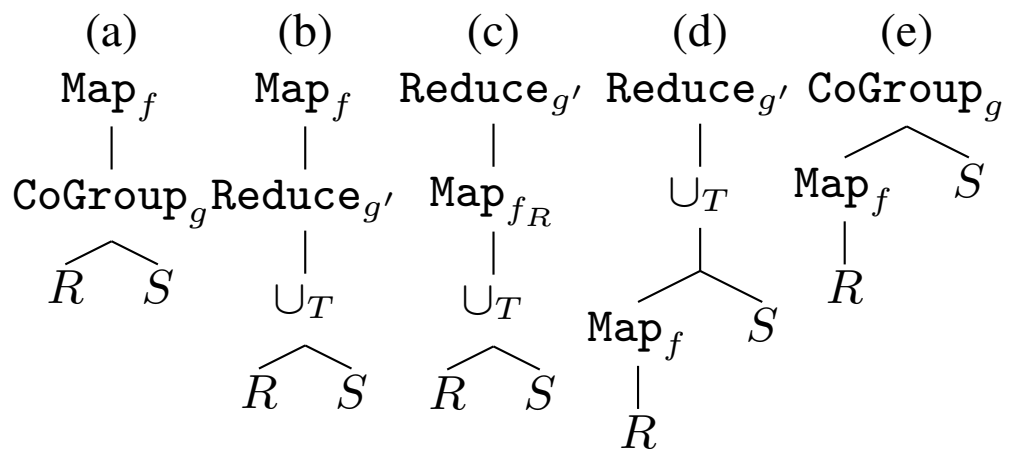

Step (a) → (b) replaces CoGroup with its Reduce equivalent. Step (b) → (c) uses the conditions of Theorem 2 to reorder the Map and Reduce operators and transforms Map's UDF $f$ to $f_R$. Step (c) → (d) pushes the Map operator under the tagged union by reversing the previous transformation. Finally, step (d) → (e) reconstructs the CoGroup operator and transforms $f_R$ back to $f$. We follow the same procedure to establish reordering conditions between CoGroup and other operators.

### 4.4 Possible Optimizations

We have presented the necessary and sufficient conditions to reorder every combination of PACT operators. These conditions are usually the ROC and the KGP conditions, together with some restrictions on the key of the Reduce operator.

These conditions lead to a number of possible optimizations. First, assuming a straightforward implementation of an acyclic SQL query as a PACT program, our conditions allow the full set of join and selection reorderings that RDBMS optimizers consider. Second, we allow the invariant grouping transformation [13], the most elementary form of aggregation push-down. More advanced transformations that include group-by considered by RDBMSs assume knowledge of the nature of the aggregating function, and are thus of limited applicability in settings of arbitrary UDFs as ours [14].

We do not allow reorderings that need semantic information to be established, including associative side-effects. For example, we cannot reorder two Map functions that add a constant number to the same field. In addition, the fact that we discover the necessary conditions for reordering through static code analysis poses further restrictions to the possible optimizations (see Section 5 for details).

## 5. DISCOVERING PROPERTIES VIA CODE ANALYSIS

The reordering proofs presented in Section 4 assume knowledge of a global record, read and write sets for each operator, as well as bounds on the output cardinality. In this section, we briefly sketch our solution to estimating read and write sets, as well as creating a global record, via static code analysis. We omit the details for emit cardinalities, which can be estimated by traversing the control flow graph of a UDF.

Our solution relies on a static code analysis (SCA) framework that analyzes the Java bytecode of a UDF. We assume that the framework provides a control flow graph and two data structures that are obtained by a data flow analysis: A use-definition chain USE-DEF($l$, `$t`) of a statement $l$ and variable `$t` is a list of all possible definitions of variable `$t` that reach $l$ without being overridden by other definitions. A definition-use chain DEF-USE($l$, `$t`) is a list of all uses of variable `$t` defined in statement $l$.

For the remainder of this section, we assume that the UDF code is formatted as typed three-address code [5]. The possible statements in three-address code are definitions of a local (e. g., `int i`) or a temporary (e. g., `int $t`) variable, assignment (e. g., `$t:=3`), branching (e. g., `if ($t<3) goto label`), as well as basic arithmetic and function calls. In addition, we assume the existence of an attribute type, `Attribute`, as well as record types `InputRecord`, and `OutputRecord`, and a set of functions that operate on these types. These functions constitute essentially the assumed record API, which is exposed to the programmer of PACT programs, and they are gradually introduced in the course of this section.

We estimate the read set $\mathsf{R}_f$ of an operator by scanning its UDF's code for statements of the form `l:$t:=getField($ir,n)`. Statement `l` stores the $n$-th field of the (parameter variable) input record `$ir` to the temporary variable `$t`. We assume that this is the only record API function to access a particular field of an input record. We further assume that integer `n` is statically computable. Recall that the $n$-th field of the input $I$ corresponds to attribute $\alpha(I, n)$ of the global record. We then look up all uses of the temporary variable `$t` in the code using the data structure DEF-USE(`$t`). If such uses exist, then we add the attribute $\alpha(I, n)$ to $\mathsf{R}_f$.

Estimating the write set $\mathsf{W}_f$ of an operator is more challenging than read set estimation since also implicit modifications must be taken into account. Our record API provides two constructors to create an output record `$or`. First, a copy constructor `$or=new OutputRecord($ir)` to copy an input record `$ir`. Second, the default constructor `$or=new OutputRecord()` to create a new and empty output record `$or`. The subtle difference is that the first constructor implicitly copies all attributes of the input record (*Implicit Copy*) while the second method implicitly projects all attributes (*Implicit Projection*). In addition, the API provides methods to explicitly copy, project, modify, and add single attributes to output records. Therefore, the code analysis method to estimate write sets must identify whether a user function implicitly copies or projects, and estimate a complementary set of explicitly projected or copied attributes. In addition, a set of modified and added attributes must be derived.

In order to identify the implicit operation and the attribute sets required for the write set estimation, we start by collecting all statements of the form `e:emit($or)` which emit the output record `$or`. We track the origin of `$or` and can safely identify the implicit operation by identifying the constructor call. If both constructors are used in different code paths, implicit projection is the safe choice. Subsequently, the remaining attribute sets are estimated by collecting all statements `s:setField($or,n,$t)`, that set the $n$-th field of output record `$or` to the value of the temporary variable `$t`. Explicit projections can be identified if `$t` is `null`. Explicit copies require that `$t` was previously set by `l:$t:=getField($ir,n)`. This can be easily detected by looking at USE-DEF(`$t`). In all other cases, statement `s` defines an explicit modification operation and is added to the appropriate set. Note that it is always safe to consider `s` as an explicit modification. Our implementation includes an additional record constructor `$o=new OutputRecord($i1,$i2)` that concatenates two input records in order to support efficient binary UDFs. This constructor yields implicit copy operations for both input records. By looking at all statements `s`, we can also keep track of the global record. A new

attribute $\alpha(O, n)$ is added to the global record if integer n is larger than #I, the number of attributes in the input $I$.

The most important property of any method that relies on static code analysis is to guarantee *safety*. In our setting, safety is defined as follows: Our analysis algorithm creates a set of properties, which in turn lead to a certain set of possible reorderings. These reorderings result in a set of plans $P'$ equivalent to the initial plan $P$. Our method is safe if $P'$ and $P$ produce the same query result for every possible input $I$.

We guarantee safety through conservatism. In particular, we guarantee that the properties discovered by our static code analysis algorithm are *supersets* of the true properties of any execution of the program for any collection of input data sets. We omit the proofs due to lack of space. The main intuition is that we consider all possible execution paths of operators, and we add an attribute to the global record, and read and write set of an operator when in doubt. Since the discovered properties are supersets of the real properties, they cause additional conflicts (see Section 4) leading to a subset of the valid reorderings, and thus to a subset of the true equivalent alternative plans.

## 6. PLAN ENUMERATION

In this section, we present an algorithm that, for a given data flow, enumerates all data flows that can be derived by valid pairwise reorderings of operators. The algorithm differs significantly from the well-known enumeration algorithms used in traditional relational database optimizers, namely enumeration via top-down branch-and-bound [19,21] or bottom-up dynamic programming [27, 29]. This is due to the difference in the algorithm input. Traditional relational optimizers operate on algebraic expressions on which heuristics such as selection and projection push-down can be applied and from which data structures such as join graphs can be derived. In contrast, our enumeration algorithm is called with a specific data flow instance from which all valid reordered data flows must be generated. In the presented version, the algorithm is restricted to tree-shaped data flows, i. e., an operator may only have a single ancestor.

Algorithm 1 provides pseudocode for enumerating all valid alternatives for a given data flow. The algorithm is based on recursive calls to enumerate alternatives for sub-flows and the exchange of two neighboring operators. In the listing, data flows and sets of data flows are denoted with capitalized names while operators and sets of operators have lowercased names. The functions `getRoot`($D$) and `rmRoot`($D$) return or remove the root of the data flow $D$, while `addRoot`($D$, $r$) appends $r$ as root of $D$ and `setRoot`($D$, $r$) replaces $D$'s root with $r$. For ease of exposition, the algorithm as shown handles data flows with single-input operators only. However, it can be easily extended to deal with non-unary operators, and our implementation can, in fact, handle binary operators.

We discuss the algorithm and argue that it computes all valid reordered data flows with the help of an example data flow $D = [\texttt{Src} \rightarrow \texttt{Map}_1 \rightarrow \texttt{Map}_2 \rightarrow \texttt{Map}_3]$. The flow consists of a data source `Src` and three Map operators with $\texttt{Map}_3$ being the root. We assume that all Map operator pairs can be reordered except for $\texttt{Map}_2$ and $\texttt{Map}_3$. The algorithm starts by recursively enumerating all reordered alternatives $Alts_{-r}$ for $D_{-r}$, which is the input data flow $D$ minus the root operator $r$ ($\texttt{Map}_3$) (Line 18):

$$Alts_{-\texttt{Map}_3} = \texttt{Enum-Alternatives}([\texttt{Src} \rightarrow \texttt{Map}_1 \rightarrow \texttt{Map}_2]) = \{[\texttt{Src} \rightarrow \texttt{Map}_1 \rightarrow \texttt{Map}_2], [\texttt{Src} \rightarrow \texttt{Map}_2 \rightarrow \texttt{Map}_1]\}$$

The result of the first recursive call $Alts_{-r}$ is used for two purposes. First, to enumerate a subset of the result $Alts$, namely all

**Algorithm 1** Enumeration of Alternative Data Flows.

1: **function** ENUM-ALTERNATIVES($D$)
2: **input:** data flow $D$
3: **output:** all possible data flows derived by reordering of $D$
4: $Alts = mTab$.get(`getMTabKey`($D$)) // check memoTable
5: **if** ($Alts \neq \emptyset$) **then**
6: **return** $Alts$
7: $r$ = `getRoot`($D$) // get root $r$ of $D$
8: **if** ($r$ is data source) **then**
9: $Alts = \{r\}$
10: **else if** ($r$ is data sink) **then**
11: $D_{-r}$ = `rmRoot`($D$)
12: $Alts_{-r}$ = `Enum-Alternatives`($D_{-r}$)
13: **for** ($A_{-r} \in Alts_{-r}$) **do** // add $r$ to each $A_{-r}$
14: $Alts = Alts \cup \{$`addRoot`$(A_{-r}, r)\}$
15: **else if** ($r$ is single-input operator) **then**
16: $cand = \emptyset$
17: $D_{-r}$ = `rmRoot`($D$)
18: $Alts_{-r}$ = `Enum-Alternatives`($D_{-r}$)
19: **for** ($A_{-r} \in Alts_{-r}$) **do**
20: $s$ = `getRoot`($A_{-r}$) // get candidate root $s$
21: $Alts = Alts \cup \{$`addRoot`$(A_{-r}, r)\}$ // add $r$ to $A_{-r}$
22: **if** ($s \notin cand \wedge$ `reorderable`$(r, s)$) **then**
23: $cand = cand \cup \{s\}$ // enum candidate $s$ only once
24: $D_{-s}$ = `setRoot`($A_{-r}, r$) // replace $s$ by $r$
25: $Alts_{-s}$ = `Enum-Alternatives`($D_{-s}$)
26: **for** $A_{-s} \in Alts_{-s}$ **do** // append $s$ to each $A_{-s}$
27: $Alts = Alts \cup \{$`addRoot`$(A_{-s}, s)\}$
28: $mTab$.put(`getMTabKey`($D$), $Alts$)
29: **return** $Alts$

reordered flows with the original root $r$. This is done by simply appending the root $r$ ($\texttt{Map}_3$) to each computed alternative $A_{-r} \in Alts_{-r}$ (Line 21):

$$Alts = \{[\texttt{Src} \rightarrow \texttt{Map}_1 \rightarrow \texttt{Map}_2 \rightarrow \texttt{Map}_3]\} \cup \{[\texttt{Src} \rightarrow \texttt{Map}_2 \rightarrow \texttt{Map}_1 \rightarrow \texttt{Map}_3]\}$$

Second, $Alts_{-r}$ is used to retrieve candidate root operators $s$ that can be reordered with $r$. For each root $s$ of the computed alternatives $A_{-r} \in Alts_{-r}$, the algorithm checks whether it can be reordered with the original root $r$ ($\texttt{Map}_3$) by calling the Boolean function `reorderable`$(r, s)$ (Line 22). In our example, this is only true for $s = \texttt{Map}_1$ and $r = \texttt{Map}_3$ since $\texttt{Map}_3$ and $\texttt{Map}_2$ cannot be reordered. Therefore, $\texttt{Map}_3$ replaces $\texttt{Map}_1$ as root of $A_{-r} = [\texttt{Src} \rightarrow \texttt{Map}_2 \rightarrow \texttt{Map}_1]$ , i. e., $r$ is pushed down to data flow $D_{-s}$ (Line 24):

$$D_{-\texttt{Map}_1} = [\texttt{Src} \rightarrow \texttt{Map}_2 \rightarrow \texttt{Map}_3]$$

The successive recursive call `Enum-Alternatives`$(D_{-s})$ enumerates all valid reorderings for the $D_{-s}$ (Line 25):

$$Alts_{-\texttt{Map}_1} = \texttt{Enum-Alternatives}([\texttt{Src} \rightarrow \texttt{Map}_2 \rightarrow \texttt{Map}_3]) = \{[\texttt{Src} \rightarrow \texttt{Map}_2 \rightarrow \texttt{Map}_3]\}$$

The result set $Alts$ is amended by all valid reorderings that have $s$ as root. This is done by simply appending $s$ to all reordered flows $A_{-s} \in Alts_{-s}$ (Line 27):

$$Alts = Alts \cup \{[\texttt{Src} \rightarrow \texttt{Map}_2 \rightarrow \texttt{Map}_3 \rightarrow \texttt{Map}_1]\}$$

Finally, all computed alternatives $Alts$ are returned:

$$Alts = \{[\texttt{Src} \rightarrow \texttt{Map}_1 \rightarrow \texttt{Map}_2 \rightarrow \texttt{Map}_3], [\texttt{Src} \rightarrow \texttt{Map}_2 \rightarrow \texttt{Map}_1 \rightarrow \texttt{Map}_3], [\texttt{Src} \rightarrow \texttt{Map}_2 \rightarrow \texttt{Map}_3 \rightarrow \texttt{Map}_1]\}$$

In order to avoid duplicate enumerations, the algorithm may only descent once into recursion for each distinct root candidate $s$ (Lines

16, 22, 23). The use of a memo table reduces the number of recursive descents and improves the runtime (Lines 4, 28).

The enumeration algorithm can also be easily integrated with a Volcano-style physical optimizer using interesting properties as described in [7, 21]. Instead of computing and returning all valid reordered data flows, the `Enum-Alternatives()` function can be adapted to compute the least expensive physical execution plan for each interesting property. Additionally, the algorithm must take care that at least one plan for each possible root node $s$ of a sub-flow is returned, in order to enumerate all possible reorderings. Physical execution plans are generated by recursively computing the least expensive execution plans for sub-flows, choosing local and shipping strategies only for the root node, and connecting it to the sub-plan. Interesting properties can be tracked during recursive descent and be used to enumerate physical execution plans for sub-flows. By integrating cost-based physical optimization in the enumeration algorithm, the principle of optimality can be exploited which effectively reduces the number of enumerated alternatives.

In contrast to optimization of relational queries, our approach for enumerating reordered data flows is limited by the choice of the initial data flow. For some queries, such as queries that include circular join graphs, the initial data flow already implies a plan decision that cannot be changed by reordering operators.

## 7. EVALUATION

We implemented a prototype to evaluate our approach for data flow optimization. The prototype is based on a pre-release snapshot of the next version of the Stratosphere system which is available as open source [4]. Furthermore, we implemented data processing tasks from different domains as PACT programs to experimentally evaluate and validate our approach. The domains include relational OLAP, as well as weblog clickstream processing and biomedical text mining. Our experimental evaluation covers the following aspects. First, we assess the optimization potential for parallel data flows. Second, we evaluate the plan space enumerated by our optimizer. Third, we discuss the overhead of the plan enumeration algorithm. Finally, we verify that static code analysis can be used to derive the necessary properties for reordering UDFs.

We start discussing our prototypical implementation and present the PACT programs used for evaluation before we show and discuss experimental results.

### 7.1 Experimental Setup

The existing optimizer of Stratosphere performs cost-based physical optimization as known from parallel relational optimizers, i. e., it selects data shipping and execution strategies such as broadcasting and hybrid-hash joins for a given data flow [7]. The cost model is a combination of network IO, disk IO, and CPU costs of UDF calls. For result size and cost estimations, the optimizer relies on hints such as *"Average Number of Records Emitted per UDF Call"*, *"CPU Cost per UDF Call"*, and *"Number of Distinct Values per Key-Set"*. These can be provided by the user, a language compiler (e. g., Hive or Pig), or obtained by runtime profiling.

In order to implement our prototype, we adapted the optimization process of Stratosphere's optimizer in the following ways. Prior to plan enumeration, the optimizer obtains information about the UDFs which is required to reason about reorderability of operators. This information can be provided by manually attached annotations or derived by an SCA component. Our SCA component is based on the Soot framework [3], which provides all features required by our code analysis technique (see Section 5). It does also take care of establishing the global record. After the information has been obtained, all valid alternative data flows are computed using the enumeration algorithm presented in Section 6. The existing cost-based optimizer [7] is called for each alternative to choose shipping and local strategies and compute a cost estimate. Finally, the cheapest plan is selected and returned for execution.

We perform our experiments on a cluster of four machines each being equipped with two Intel Xeon E5530 Quadcore CPUs, 48 GB RAM, and ten 250 GB disks for data bundled in a RAID5. The machines are connected with 1 GBit Ethernet and run Linux (Ubuntu Server 10.04.3 LTS), Sun Java 6, and HDFS 0.20.2. We execute all tasks with a degree of parallelization of 32.

### 7.2 Evaluation Programs

We evaluate our approach using four tasks from different domains. Algebraic optimization of relational queries is best known from relational DBMS but also applied in the context of parallel data flow systems by higher-level languages such as Hive [31], SCOPE [12], and Tenzing [16]. In order to show the effectiveness of our approach, we implemented two queries of the TPC-H benchmark for evaluation. Parallel data flow engines are commonly used for non-relational tasks. We show the applicability of data flow optimization for such domains by providing two non-relational tasks, namely biomedical text mining and weblog clickstream processing. All four tasks are implemented as handcrafted PACT data flows. In this section, we shortly present all tasks and their implementations.

**Relational OLAP:** We implemented slightly modified variants of queries 7 (where we reduced the selectivity of the `shipdate` filter and removed the final sorting) and 15 (where we removed the filter on `total_revenue`) from the TPC-H benchmark to cover relational analytical tasks. For our experiments, we run both queries on a 400 GB TPC-H data set. Query 7 applies a local predicate on the lineitems relation, joins six relations with circular-connected join predicates, and finally performs a grouping with sum aggregation. Figure 2(a) shows our PACT implementation. All joins are implemented as Match operators except the join with the disjunctive join predicate ($\mathtt{nation}_1 \bowtie \mathtt{nation}_2$), which is implemented as a filtering Map operator. Grouping and sum aggregation are done by a Reduce operator.

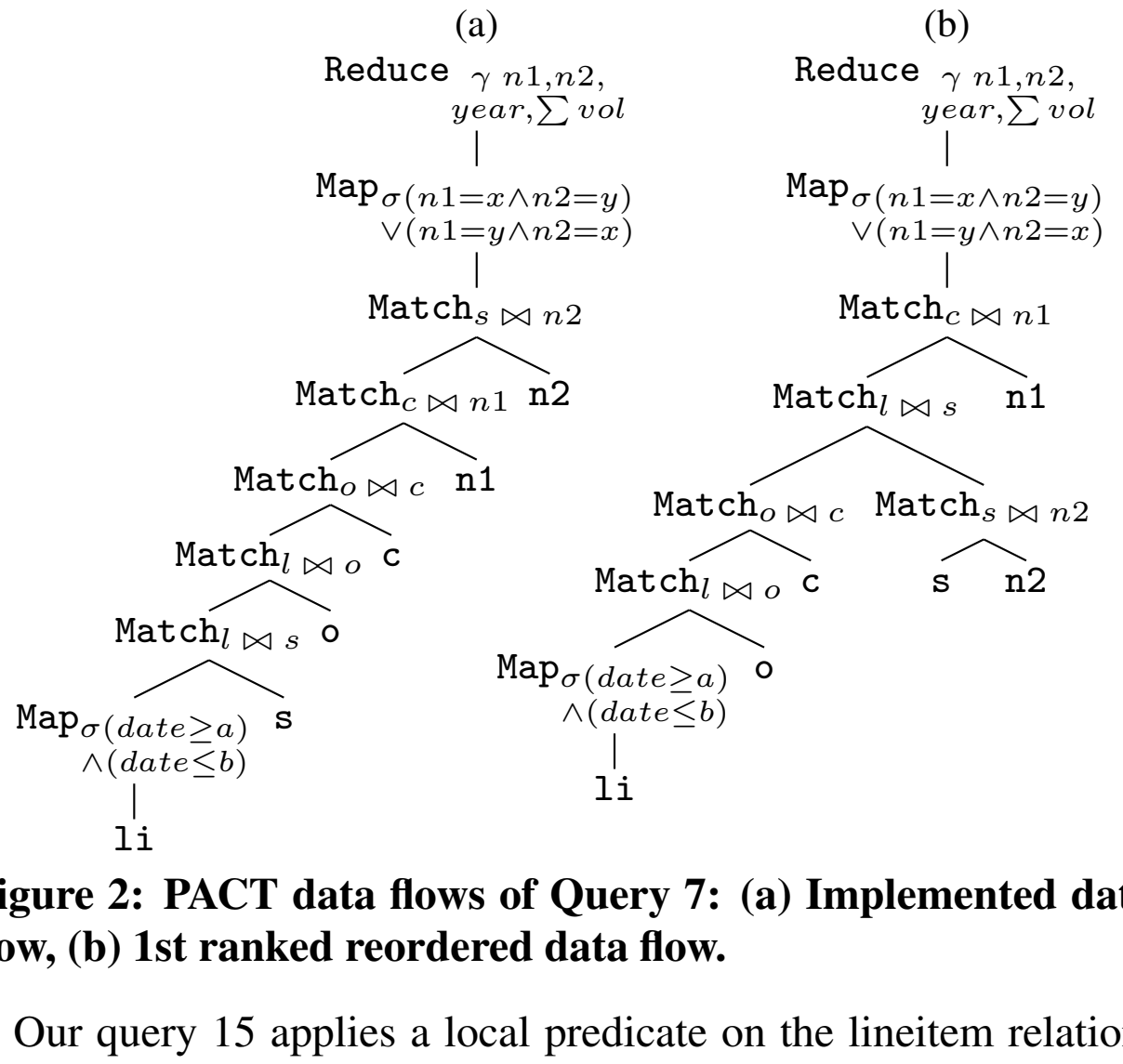

**Figure 2: PACT data flows of Query 7: (a) Implemented data flow, (b) 1st ranked reordered data flow.**

Our query 15 applies a local predicate on the lineitem relation, joins it with the supplier table, and groups and aggregates to compute the final result. We implemented the local predicates as a Map, the join as a Match, and the grouping and aggregation as a Reduce operator (see Figure 3(a)). Note that the join predicate and the grouping clause reference the same attribute (s_key). As shown in Section 4.3.2, this is a necessary condition to reorder a Match and a Reduce operator.

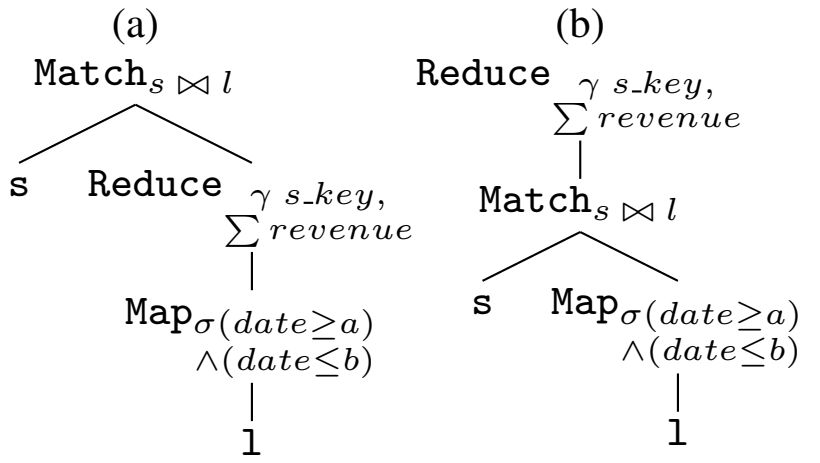

**Figure 3: PACT data flows of Query 15: (a) Implemented data flow, (b) Data flow with Match before Reduce.**

**Text Mining:** We implemented a text mining task that detects relationships between genes and drugs described in biomedical text corpora. The data flow is a pipeline of Map operators which extract entities and relationships by applying several natural language processing (NLP) algorithms to the input. Our program takes a text collection as input and performs some linguistic preprocessing, e. g., tokenization and part-of-speech tagging on the input, to enable entity and relation extraction. In order to reduce intermediate result set sizes, each entity or relation extraction component also works as a filter by forwarding only those records that actually contain a gene, drug, or relation mention. Most NLP components are very compute-intensive since they often call third-party machine-learning or automaton-based components to enable the extraction process. Furthermore, most components have dependencies on other components to be executed in advance. These dependencies limit the set of valid reordered data flows. Optimization potential arises from different filter selectivities and varying execution costs for the text mining components. We execute the text mining data flow on a 425 MB subset of the PubMed data set.

**Clickstream Processing:** Weblog processing is a common example of non-relational data flows [20]. We implemented a task that processes web shop clickstream data (see Figure 4(a)). The task extracts click sessions that lead to buy actions and augments them with detailed user information. Such tasks are common preprocessing steps for data mining algorithms. In our scenario, a clickstream entry contains an IP address, a timestamp, and a visited URL. The URL encodes a session id, and the performed user action. The first Reduce operator filters on sessions that include at least one buy action. The successive Reduce operator condenses a session into a single record. The following Match operator joins by session id with a relation that resolves session ids to ids of logged in users, thereby selecting only sessions with logged in users. Finally, a second Match operator appends detailed user information by joining on the user id. For our experiments, we ran the task on 430 GB click, 13.8 GB login, and 9.2 GB user_info data.

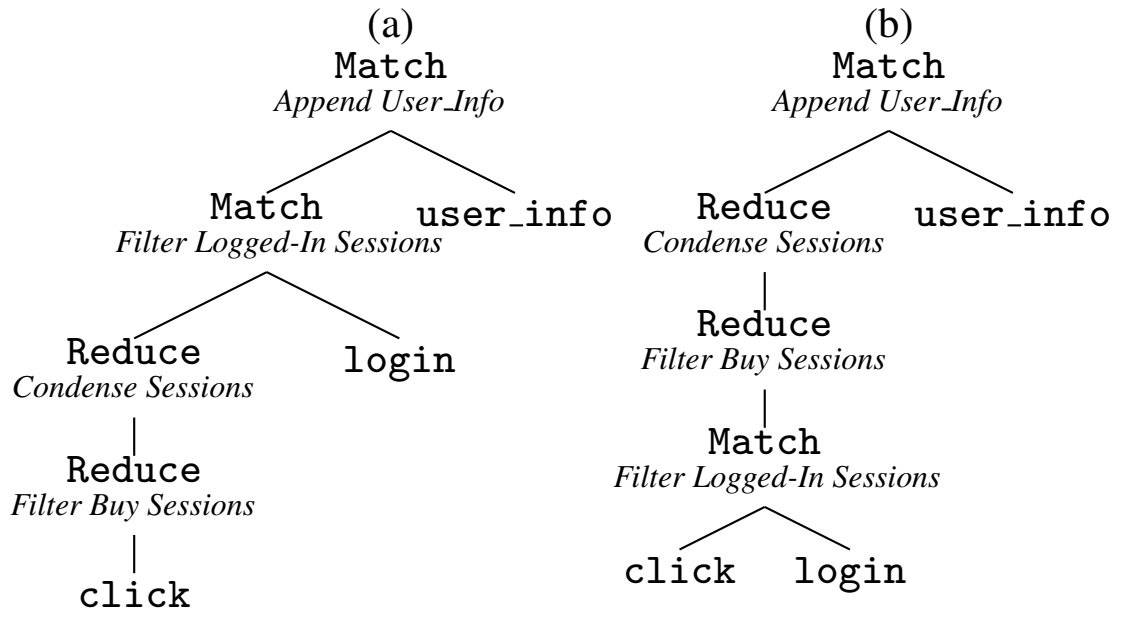

**Figure 4: PACT data flows of clickstream processing task: (a) Implemented data flow, (b) 1st ranked reordered data flow.**

## 7.3 Experiments

**Optimization Potential:** Query optimization as performed by modern relational DBMSs has the potential to improve query execution time by orders of magnitude. Our first set of experiments assesses the potential of our generic data flow optimization technique. We enumerate all possible data flows for a given PACT program. Each reordered alternative is fed into the physical optimizer where shipping and local execution strategies are enumerated, and cost estimates are obtained. We sort the resulting plans in ascending order by their estimated costs and assign a rank to each plan that corresponds to its position in the list. We pick ten plans in regular rank intervals from the list and execute them. For each executed plan, we plot the cost estimate of the optimizer and the actual runtime (averaged over three runs), both normalized by the lowest estimated costs and averaged runtime respectively.

Figure 5 shows the results for TPC-H query 7. The enumeration algorithm explored a space of 2518 alternative plans. We see that the plan with the least estimated costs provides also the least execution time with an absolute runtime of roughly 6 minutes (see Figure 2(b) for this plan). The last ranked plan is slower by a factor of 7 and requires the most time for execution (about 45 minutes).

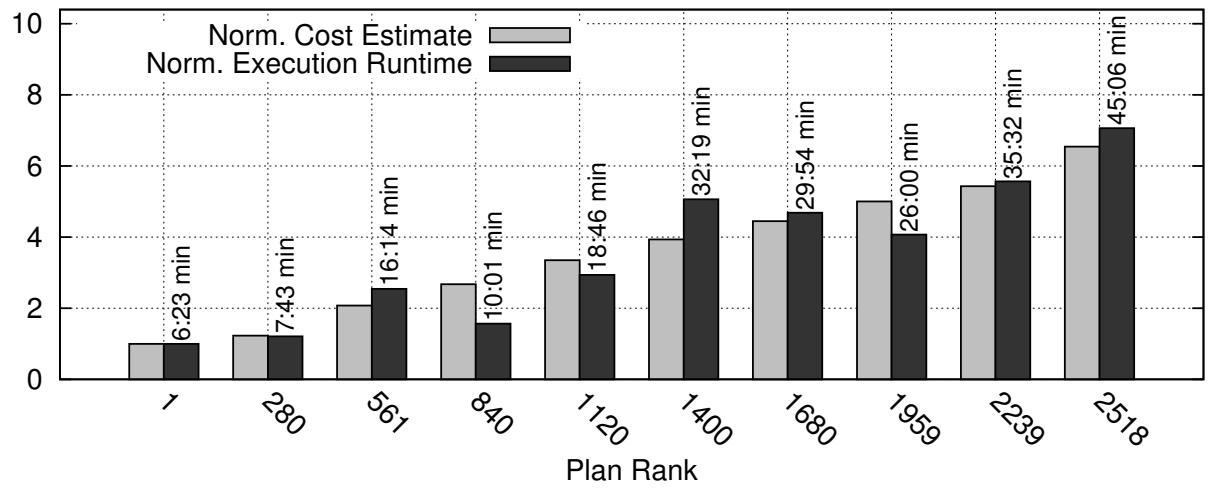

**Figure 5: Normalized cost estimates and execution runtime for 10 regularly picked execution plans of the TPC-H query 7.**

Figure 6 shows the estimated costs and runtimes for selected plans of the text mining task. The best plan (according to estimated costs) outperforms the worst by almost an order of magnitude.

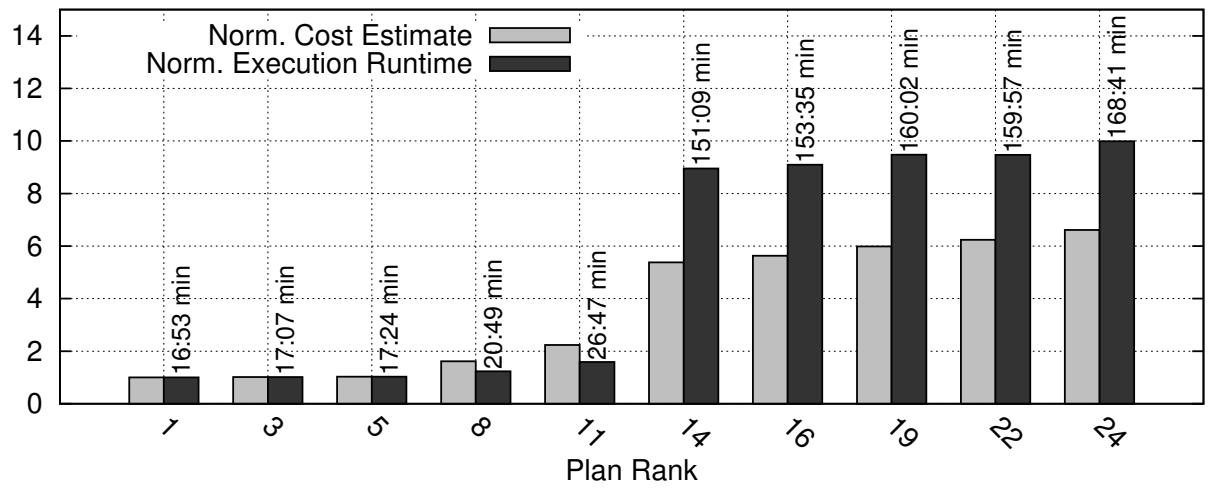

**Figure 6: Normalized cost estimates and execution runtime for 10 regularly picked execution plans of the text mining job.**

Our experiments show that reordering of data flows can lead to significant performance improvements. Due to the observation that in general execution plans with higher cost estimates require more time for execution, we can also approve the validity of the optimizer's cost model. We note that Stratosphere does not support indexes, columnar layouts, or materialized views yet. Therefore, all execution plans result in full scans of all included data sets, which limits the achievable runtime improvements.

**Plan Enumeration Space:** We continue discussing the plan enumeration space with TPC-H query 15. Our implementation is based on a Map, a Reduce, and a Match operator (see Figure 3). We can exchange Match and Reduce since the ROC condition is fulfilled, Match preserves the group cardinality because it is a PK-FK join, and Reduce groups on the match key (s_key). This is essentially an aggregation push-up rewrite that could also be applied by a relational optimizer. Besides the changed order of Reduce and Match, the rewrite also leads to different physical plan choices.

For the data flow with Reduce being the input of Match (Figure 3(a)), the physical optimizer chooses to partition the input of Reduce and establish the groups by sorting. The grouped and aggregated result is locally forwarded into the Match operator and used to build a hash table. Since Match operates on the same key as Reduce, the partitioning property remains and can be reused. To compute the final result, the supplier relation is also partitioned, shipped to the Match operator, and probed against the hash table. In fact, the optimizer could also choose to reuse the sorting of Reduce and perform a sort-merge join for Match. However, this would require to sort the supplier relation.

The alternative data flow with Match being the input of Reduce (Figure 3(b)) is executed using a different shipping strategy. In this case, Match's lineitem input is much larger than the supplier input, since it has not been aggregated as in the previous case. Therefore, the optimizer decides to broadcast the much smaller supplier input to all parallel instances of Match and build a hash table from it. The lineitem side is locally forwarded and probed against the hash table. The result is partitioned and shipped to Reduce which groups by sorting and computes the final result.

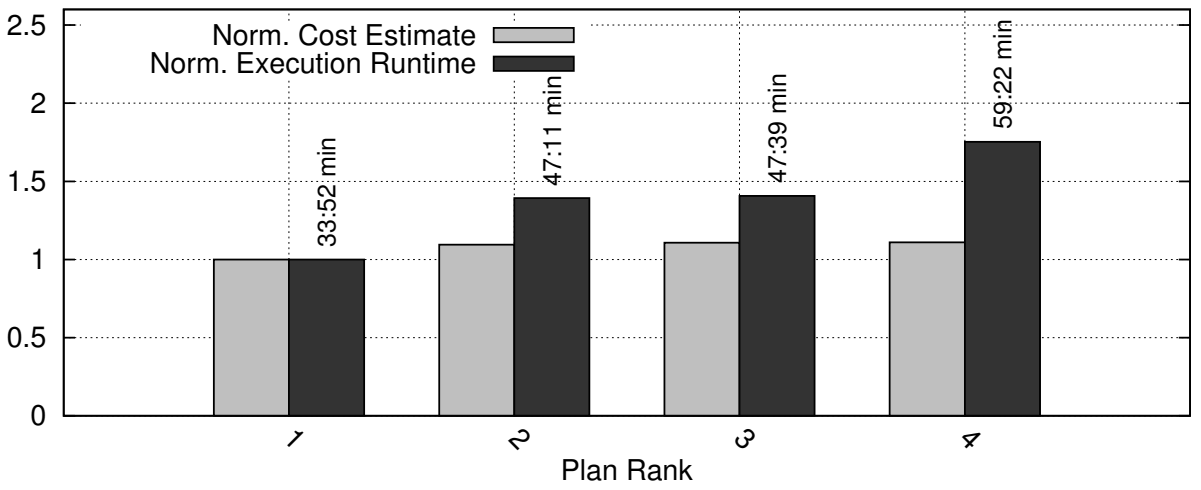

**Figure 7: Normalized cost estimates and execution runtime for all 4 execution plans of the clickstream processing job.**

As previously stated, our optimizer is also able to reorder non-relational operators. Figure 4 shows (a) the implemented PACT program and (b) the data flow chosen by the optimizer for the clickstream processing task. Both Reduce operators are non-relational operators. The *"Filter Buy Sessions"* UDF is called with all click records of a session and checks whether at least one click performs a buy action. In that case, all click records are forwarded, otherwise none. The subsequent *"Condense Sessions"* UDF collects all clicks of a session, merges them into a single record and forwards it. Comparing the best performing and the implemented data flow, we see that the optimizer pushed the selective join (*"Filter Logged-In Sessions"*) below both non-relational Reduce operations. We are not aware of a data processing system that is able to perform similar optimizations. Figure 7 gives the estimated costs and runtimes of all four execution plans. The best performing plan beats the implemented data flow (Rank 3) by a factor of 1.4 or 13:47 minutes.

Our optimizer explores large fractions of the search space that conventional relational optimizers cover, including bushy join orders (Figure 2), pushed aggregations (Figure 3), and reasoning about interesting properties [7]. Furthermore, we show that our approach enables optimizations that are not supported by any current data analysis system we are aware of (Figure 4).

**Enumeration Time:** Our enumeration algorithm is facing the same problem of exponential search space sizes as relational optimizers. As previously discussed, our prototypical implementation first enumerates all valid reordered data flows and subsequently calls the physical optimizer for each candidate. This implementation does not permit cost-based search space pruning and it is not tailored towards efficient plan enumeration. In Section 6 we gave an intuition how the enumeration algorithm could integrated with physical optimization. An important part of our future research is to leverage well-known search space pruning techniques and benchmark the overhead of our query optimizer. For all queries presented so far, which represent typical data analysis tasks, plan enumeration took less than 1654 ms using our naive implementation. The overhead of performing the static code analysis is virtually zero.

| PACT Task | Enumerated Orders with Manual Annotation | Enumerated Orders with SCA |
|---|---|---|
| Clickstream | 4 | 3 (75%) |
| TPC-H Q7 | 2518 | 2518 (100%) |
| TPC-H Q15 | 4 | 4 (100%) |
| Text Mining | 24 | 24 (100%) |

**Table 1: Comparing number of reordered alternatives for manually annotated and automatically derived read and write sets.**

**Feasibility of Static Code Analysis:** We evaluate the feasibility of static code analysis to determine read and write sets of UDFs. For this purpose, we compare the number of reordered alternative data flows that were enumerated based on read and write sets which were manually annotated and automatically derived using static code analysis. Table 1 gives the results for all presented evaluation tasks. The information extracted by our prototypical implementation of the SCA component enables the optimizer to enumerate almost all valid plans for our four evaluation data flows. The current implementation is restricted to information that is available at UDF compile time and can be easily accessed such as field accesses with literals and final variables. This can be extended to more exhaustive control flow tracking and incorporation of job configurations which are only available at optimization time.

# 8. RELATED WORK

We are not aware of any work that aims to optimize data flows with unknown operator semantics by reordering. The work most relevant to ours is Manimal [26]. Manimal applies static code analysis to MapReduce programs to identify relational-style selections and projections. An optimizer selects from available $B^+$-tree indexes and decides on the use of delta-compression. Manimal's optimizations are orthogonal to ours (operator reordering), and would constitute valuable additions to our system.

Optimization of user-defined predicates was discussed in the context of extensible RDBMSs [15, 23]. This line of work only considered UDFs with the semantics of relational selection operators. Therefore, the challenge was not to identify when reordering is possible, but when it is beneficial.

A number of recent approaches consider optimization in the context of translating a higher-level algebraic specification to a data flow. Higher-level specifications include AQL [9], Pig [28], Jaql [10], Hive [31], Tenzing [16], DryadLINQ [33], and SCOPE [12]. The target parallel data flow platforms include MapReduce [17], Dryad [25], and Hyracks [11]. In contrast to these approaches, our work applies optimization directly to data flows without knowledge of the operator's algebraic properties. We view the two approaches as complementary; while we show that some optimizations can be done at the data flow level, thus making a data flow engine able to seamlessly handle multiple data and programming models, other optimizations are semantic in nature and can only be done at a higher level. We note that higher-level language translators can enrich data flows with reordering information based on the operators semantics, hence enabling the unified optimization of operator order and physical optimization at the data flow abstraction.

The Starfish project applies cost-based optimization to MapReduce programs [24]. In contrast to our work, Starfish does neither inspect or optimize the program itself. Instead, it uses runtime profiling and cost-based optimization to generate well-performing job configurations for Hadoop MapReduce jobs.

Finally, we draw inspiration from the Ferry project [22]. Ferry follows an algebraic approach to push data processing instructions from the application into the DBMS by translating general-purpose (application) code to SQL queries.

# 9. CONCLUSIONS AND FUTURE WORK

We propose and address the problem of optimizing data flows that consist of black box user-defined functions written in an imperative language. In this setting, the algebraic properties of the operators of the data flow are unknown, and must be discovered. Our key insight is that a handful of properties, which can be discovered using static code analysis, suffice to establish many optimizations known from relational algebra, including filter and join reordering, and some forms of aggregation push-down. We formally establish reordering conditions, show how to estimate the desired properties via static code analysis, and present a plan enumeration algorithm.

We have prototyped our solution in the Stratosphere system. Our experimental results show that our approach is able to reorder relational and non-relational data flows, leading to runtime improvements of up to an order of magnitude. Moreover, we demonstrate that our approach is able to perform optimizations which algebraic optimizers are not capable of. Our experiments attest, that our static code analyzer successfully extracts properties from black box UDFs that are required for reordering them.

We identify several avenues for future research. While this work explores which reorderings of data flows are possible, we plan to identify which reorderings are beneficial. This will include estimating the selectivity and execution cost of black box operators. Furthermore, we plan to investigate a wider range of optimizations including managing attribute projections globally in a plan, optimizations that take into account some semantic information of operators, and *intrusive* optimizations that change the code of the user functions. The latter could include, dissecting an operator into independent components that can be then individually reordered. We plan to exploit the state of the art of semantic program analysis to gain more information about the internals of the operator UDFs.

## Acknowledgments

This research was funded by the German Research Foundation under grant "FOR 1036: Stratosphere-Information Management on the Cloud." We thank Stephan Ewen, Johann-Christoph Freytag, Ulf Leser, Volker Markl, and Matthias Ringwald for valuable discussions and support.

# 10. REFERENCES

[1] `http://public.web.cern.ch/public/en/LHC/Computing-en.html`.

[2] `http://www.greenplum.com/technology/mapreduce`.

[3] `http://www.sable.mcgill.ca/soot/`.

[4] `http://stratosphere.eu`.

[5] A. V. Aho, M. S. Lam, R. Sethi, and J. D. Ullman. *Compilers: Principles, Techniques and Tools*. Pearson, 2006.

[6] M. Baker. Next-generation sequencing: Adjusting to data overload. *Nature Methods*, 7(7):495–499, 2010.

[7] D. Battré, S. Ewen, F. Hueske, O. Kao, V. Markl, and D. Warneke. Nephele/PACTs: A programming model and execution framework for web-scale analytical processing. In *SoCC*, pp. 119–130, 2010.

[8] J. Becla, A. Hanushevsky, S. Nikolaev, G. Abdulla, A. S. Szalay, M. A. Nieto-Santisteban, A. Thakar, and J. Gray. Designing a multi-petabyte database for LSST. *CoRR*, abs/cs/0604112, 2006.

[9] A. Behm, V. R. Borkar, M. J. Carey, R. Grover, C. Li, N. Onose, R. Vernica, A. Deutsch, Y. Papakonstantinou, and V. J. Tsotras. ASTERIX: Towards a scalable, semistructured data platform for evolving-world models. *Distributed and Parallel Databases*, 29(3):185–216, 2011.

[10] K. S. Beyer, V. Ercegovac, R. Gemulla, A. Balmin, M. Y. Eltabakh, C.-C. Kanne, F. Özcan, and E. J. Shekita. Jaql: A scripting language for large scale semistructured data analysis. *PVLDB*, 4(12):1272–1283, 2011.

[11] V. R. Borkar, M. J. Carey, R. Grover, N. Onose, and R. Vernica. Hyracks: A flexible and extensible foundation for data-intensive computing. In *ICDE*, pp. 1151–1162, 2011.

[12] R. Chaiken, B. Jenkins, P.-Å. Larson, B. Ramsey, D. Shakib, S. Weaver, and J. Zhou. SCOPE: Easy and efficient parallel processing of massive data sets. *PVLDB*, 1(2):1265–1276, 2008.

[13] S. Chaudhuri and K. Shim. Including group-by in query optimization. In *VLDB*, pp. 354–366, 1994.

[14] S. Chaudhuri and K. Shim. An overview of cost-based optimization of queries with aggregates. *IEEE Data Eng. Bull.*, 18(3):3–9, 1995.

[15] S. Chaudhuri and K. Shim. Optimization of queries with user-defined predicates. *ACM TODS*, 24(2):177–228, 1999.

[16] B. Chattopadhyay, L. Lin, W. Liu, S. Mittal, P. Aragonda, V. Lychagina, Y. Kwon, and M. Wong. Tenzing A SQL Implementation On The MapReduce Framework. *PVLDB*, 4(12):1318–1327, 2011.

[17] J. Dean and S. Ghemawat. MapReduce: Simplified data processing on large clusters. In *OSDI*, pp. 137–150, 2004.

[18] J. Dean and S. Ghemawat. MapReduce: Simplified data processing on large clusters. *CACM*, 51(1):107–113, 2008.

[19] P. Fender and G. Moerkotte. A new, highly efficient, and easy to implement top-down join enumeration algorithm. In *ICDE*, pp. 864–875, 2011.

[20] E. Friedman, P. M. Pawlowski, and J. Cieslewicz. SQL/MapReduce: A practical approach to self-describing, polymorphic, and parallelizable user-defined functions. *PVLDB*, 2(2):1402–1413, 2009.

[21] G. Graefe. The Cascades framework for query optimization. *IEEE Data Eng. Bull.*, 18(3):19–29, 1995.

[22] T. Grust, M. Mayr, J. Rittinger, and T. Schreiber. Ferry: Database-supported program execution. In *SIGMOD*, pp. 1063–1066, 2009.

[23] J. M. Hellerstein. Optimization techniques for queries with expensive methods. *ACM TODS*, 23(2):113–157, 1998.

[24] H. Herodotou and S. Babu. Profiling, what-if analysis, and cost-based optimization of MapReduce programs. *PVLDB*, 4(11):1111–1122, 2011.

[25] M. Isard, M. Budiu, Y. Yu, A. Birrell, and D. Fetterly. Dryad: Distributed data-parallel programs from sequential building blocks. In *EuroSys*, pp. 59–72, 2007.

[26] E. Jahani, M. J. Cafarella, and C. Ré. Automatic optimization for MapReduce programs. *PVLDB*, 4(6):385–396, 2011.

[27] G. Moerkotte and T. Neumann. Dynamic programming strikes back. In *SIGMOD*, pp. 539–552, 2008.

[28] C. Olston, B. Reed, U. Srivastava, R. Kumar, and A. Tomkins. Pig Latin: A not-so-foreign language for data processing. In *SIGMOD*, pp. 1099–1110, 2008.

[29] P. G. Selinger, M. M. Astrahan, D. D. Chamberlin, R. A. Lorie, and T. G. Price. Access path selection in a relational database management system. In *SIGMOD*, pp. 23–34, 1979.

[30] M. Stonebraker, C. Bear, U. Çetintemel, M. Cherniack, T. Ge, N. Hachem, S. Harizopoulos, J. Lifter, J. Rogers, and S. B. Zdonik. One size fits all? Part 2: Benchmarking studies. In *CIDR*, pp. 173–184, 2007.

[31] A. Thusoo, J. S. Sarma, N. Jain, Z. Shao, P. Chakka, S. Anthony, H. Liu, P. Wyckoff, and R. Murthy. Hive - A warehousing solution over a map-reduce framework. *PVLDB*, 2(2):1626–1629, 2009.

[32] D. Warneke and O. Kao. Nephele: Efficient parallel data processing in the cloud. In *SC-MTAGS*, 2009.

[33] Y. Yu, M. Isard, D. Fetterly, M. Budiu, Ú. Erlingsson, P. K. Gunda, and J. Currey. DryadLINQ: A system for general-purpose distributed data-parallel computing using a high-level language. In *OSDI*, pp. 1–14, 2008.